# GEOS RR Lyrae Survey: Blazhko Period Measurement of Three RRab Stars—CX Lyrae, NU Aurigae, and VY Coronae Borealis

**Pierre de Ponthière**
*15 Rue Pré Mathy, Lesve, Profondeville 5170, Belgium*

**Jean–François Le Borgne**
*14, Avenue Edouard Belin, F–31400 Toulouse, France*

**F. Fumagalli**
*Calina Observatory, Carona, Switzerland*

**Franz–Josef Hambsch**
*12 Oude Bleken, Mol, 2400, Belgium*

**Tom Krajci**
*P. O. Box 1351, Cloudcroft, NM 88317*

**J.-M. Llapasset**
*66 Cours de Lassus, F–66000 Perpignan, France*

**Kenneth Menzies**
*318A Potter Road, Framingham MA 01701*

**Marco Nobile**
*via Cantonale 53, 6942 Savosa, Switzerland*

**Richard Sabo**
*2336 Trailcrest Drive, Bozeman MT 59718*

**Abstract**   We present the results of collaborative observations of three RR Lyrae stars (CX Lyr, NU Aur, and VY CrB) which have a strong Blazhko effect. This work has been initiated and performed in the framework of the GEOS RR Lyr Survey (Groupe Européen d'Observations Stellaires). From the measured light curves, we have determined the times and the magnitudes at maximum. The times of maxima have been compared to ephemerides to obtain the (O–C) values and from a period analysis of these (O–C) values, the Blazhko period is derived. The Blazhko periods of NU Aur (114.8 days) and VY CrB (32.3 days) are reported here for the first time and a more accurate period for CX Lyr (68.3 days) has been obtained. The three stars are subject to strong Blazhko effect, but this effect has different characteristics for each of them. When we compare the variations of magnitude at maximum and variations of (O–C) values with respect to the Blazhko phase, these variations are in phase, in opposition, or even in quadrature.

## 1. Introduction

The main objective of the GEOS RR Lyr Survey is to follow the variations of period and Blazhko effect of bright and well-studied RR Lyrae stars. These variations are followed in the long term with TAROT robotic telescopes (Le Borgne *et al.* 2007 and Poretti *et al.* 2008). The second objective of the survey is the observation of Blazhko effect of under-studied RR Lyrae stars. The results presented here are in keeping with this objective.

The RR Lyrae stars of Bailey type ab (RRab) are pulsating stars with a period between 0.4 and 0.7 day. Some RRab stars exhibit a phase and amplitude modulation. This phenomenon, known for a century, is called the Blazhko effect. It is recognized that this effect is still not well understood. RRab stars exhibiting the Blazhko effect appear to show a variety of characteristics. Recent continuous, high precision photometry from the Kepler satellite documents a period doubling for some RR Lyrae stars (Szabó *et al.* 2010). With our ground–based small aperture telescopes and their limited photometric accuracy, we attempt to determine the Blazhko period of neglected RRab stars. Monitoring during several years is needed to determine the Blazhko period and to characterize the Blazhko behavior. We have analyzed the variations of the magnitude at maximum and (O–C) value with respect to the Blazhko phase for three different stars (CX Lyr, NU Aur, and VY CrB).

After dark and flat field corrections with the MAXIM DL software (Diffraction Limited 2004), aperture photometry was performed using either AIP4WIN (Berry and Burnell 2001) or LESVEPHOTOMETRY (de Ponthière 2010), a custom software which also evaluates the SNR and estimates magnitude errors. No color corrections have been applied to the measured magnitudes. The times of maxima of the light curves have been evaluated with the same custom software fitting the light curve with a smoothing spline function (Reinsch 1967). We have used the ANOVA algorithm of PERANSO (Vanmunster 2007) to derive the Blazhko period from the times of maxima.

## 2. CX Lyr

The star CX Lyr is classified in the *General Catalogue of Variable Stars* (GCVS; Samus *et al.* 2011) as an RRab variable star with a period of 0.61664495 day. CX Lyr observations during the second half of 2008 (JD 2454637 to 2454783) have been previously reported by de Ponthière *et al.* (2009). During a new observation campaign from 2009 to 2011 (JD 2455041 to 2455807), we obtained forty–one new maxima.

The comparison stars used by the authors are given in Table 1. The star coordinates and magnitudes in *B* and *V* bands were obtained from the NOMAD catalogue (Zacharias *et al.* 2011). C1 was used as magnitude reference and the others as check stars. The choice of different comparison stars creates a magnitude offset due to their color differences. This offset has been evaluated by comparing the magnitudes of a common check star and taken into account. Table 2 provides the list of these new observations and Figure 1 shows the (O–C) values. For the sake of completeness, observations obtained by G. Maintz (Huebscher *et al.* 2008, 2010) and older GEOS observations are included in the table as they are used in the present analysis.

A linear regression of all available (O–C) values has provided a new pulsation period of 0.616758 day. The (O–C) values have been re–evaluated with this new pulsation period. The new elements are:

$$\text{HJD} = 2454677.5692 \pm 0.0031 + (0.6167582 \pm 0.0000031)\, E \qquad (1)$$

These values are very close to the values reported previously (de Ponthiere *et al.* 2009).

$$\text{HJD} = 2454677.5688 \pm 0.0037 + (0.61675 \pm 0.000024)\, E \qquad (2)$$

The Blazhko period was determined by a period analysis of the (O–C) values with the ANOVA algorithm. The most significant period is $68.3 \pm 0.4$ days (5.34 c/y). The periodogram presented in Figure 2 indicates other peaks at 56.6 days (6.45 c/y), 84.1 days (4.34 c/y), and 113.3 days (3.22 c/y) which are one-year sampling aliases.

There is also another peak at 136 days, that is, twice the most significant period. Data from the year 2010 (JD 2455300 to 2455500) indicate that the successive Blazhko cycles are not identical (Figure 1). The variations of successive cycles create spectral response at a multiple of the fundamental period. An (O–C) folded light curve at 136 days, would show two maxima. A similar period analysis of the magnitude at maximum with the ANOVA algorithm has provided similar conclusions.

The folded (O–C) and magnitude at maximum curves versus the Blazhko phase are given in Figure 3a and 3b. It can be seen that these two curves are nearly in phase, with the minima reached at the same Blazhko phase.

## 3. NU Aur

The star NU Aur is classified in the GCVS (Samus *et al.* 2011) as an RRab variable star with a period of 0.53941672 day and a Blazhko period of 179 days. During a first observation campaign, between December 2006 and February 2007 (JD 2454081 to 2454135), the eighteen obtained maxima clearly showed a strong Blazhko effect but did not allow a determination of the Blazhko period. The observation of seventy–five maxima resulted from a second series of observations between December 2008 and March 2011 (JD 2454752 to 2454640). The comparison stars are documented in Table 3. Star coordinates and *B* and *V* magnitudes are those found in the AAVSO's Comparison Star Database (VSD). The times of maximum and (O–C) values are given in Table 4 and Figure 4. The observations of G.Maintz have already been published by Huebscher *et al.* (2009) and those of K. Menzies published by Samolyk (2011).

A linear regression on the (O–C) values has provided the following elements:

$$\text{HJD} = 2454752.4603 \pm 0.0014 + (0.5394148 \pm 0.0000015) \text{ E} \quad (3)$$

To determine the Blazhko period we have performed a period analysis of the (O–C) values with the ANOVA algorithm. The corresponding periodogram presented in Fig. 5a indicates that the values of the four prominent peaks include:

| Period (days) | Cycles / year | Peak value |
|---|---|---|
| 114.4 | 3.192 | 49.6 |
| 170.1 | 2.148 | 56.7 |
| 227.1 | 1.600 | 47.3 |
| 339.6 | 1.074 | 53.4 |

The first two peaks (114.4 ± 1.7 and 170.1 ± 2.6 days) are an alias pair. One frequency is the alias at one cycle per year of the other. The third period (227.1) days is approximately double the first one (114.4). The period of 170.1 days is close to the value reported in the GCVS (175 days). These aliases are artifacts arising from gaps between normal 6–month observing seasons. With the Spectral Window tool in peranso, we have tried to determine which peaks are artifacts of the seasonal sampling. This algorithm calculates the pattern caused by the structure of gaps in the observations. The output of the Spectral Window is given in Figure 5b, where it can be seen that the artifact peaks are broad. The list of prominent peaks is:

| Period (days) | Cycles / year | Peak value |
|---|---|---|
| 91.1 | 4.007 | 0.08 |
| 103.4 | 3.536 | 0.06 |
| 128.8 | 2.834 | 0.07 |
| 145.2 | 2.517 | 0.16 |
| 181.4 | 2.013 | 0.12 |
| 244.2 | 1.494 | 0.28 |
| 362.2 | 1.008 | 0.60 |

This Spectral Window analysis indicates that the first peak (114.4 days) of the ANOVA analysis is not an artifact due to seasonal sampling. The second peak (170.1 days) is close to the 181.4 peak of the Spectral Window analysis and could be an artifact. The Blazhko period is probably 114.4 ± 1.7 days, but we can not eliminate the second possible period of 170.1 ± 2.6 days. More observations are needed to remove this ambiguity.

Using the adopted period of 114.4 days, the folded (O–C) and magnitude at maximum curves versus the Blazhko phase are given in Figure 6a and 6b. It can be seen that these two curves are not in phase as was the case for CX Lyr. For NU Aur star the two curves are in quadrature.

## 4. VY CrB

The star VY CrB is classified in the GCVS (Samus *et al.* 2011) as an RRab variable star. VY CrB is also designated as GSC 2576–0980 (Space Telescope Science Institute 2001). It was identified as an RRab star on photographic plates by Antipin (1996). VY CrB is herein identified as Antipin's Var 23 with a period of 0.462957 day.

We observed two maxima of VY CrB in April 2007 (JD 2454215 and 2454216) and forty–nine maxima between April 2010 and August 2011 (JD 2455302 to 2455784). The selected comparison stars are given in Table 5. Star coordinates and *B* and *V* magnitude are obtained from the NOMAD catalogue. The times of maximum and (O–C) values are given in Table 6 and Figure 7. This table also includes a previous observation obtained by A. Paschke (Agerer and Huebscher 2002).

A linear regression of the (O–C) values has provided the following elements:

$$HJD = 2455302.5032 \pm 0.0013 + (0.4629461 \pm 0.0000010)\, E \qquad (4)$$

As for the other stars, the Blazhko period has been derived with the ANOVA algorithm applied to the (O–C) values. The corresponding periodogram is given in Figure 8. The periods for the two prominent peaks are: 32.3 ± 0.1 and 64.6 ± 0.2 days, which differ by a factor of two. As for CX Lyr, the non–repetitive behavior of the Blazhko effect generates spectral response at multiples of the fundamental period. With a Blazhko period of 64.4 days we would have two (O–C) maxima per cycle, so we retained the Blazhko period value of 32.3 ± 0.1 days.

The folded (O–C) and magnitude at maximum versus the Blazhko period of 32.3 days are given in Figure 9a and 9b. It can be seen that these two curves are in phase opposition: the maximum value of (O–C) occurs when the magnitude at maximum is at its minimum value.

## 5. Blazhko behavior comparison

It is interesting to plot the magnitude at maximum versus the (O–C) values. If these quantities were varying in time as sinusoids and were in phase, the resulting graph would be a straight line in the first and third quadrants. If they were in phase opposition, the graph would be a straight line in the second and fourth quadrants and if they were in quadrature the graph would exhibit a circle. The periodical variations of magnitude at maximum and the (O–C) values are not sinusoidal, but the corresponding parametric representation will nevertheless provide useful information.

These graphs for the three stars are given in Figure 10. For the CX Lyr, the points are scattered along two segments forming a right angle but the general trend is a slope at 45 degrees indicating that (O–C) and magnitude at maximum are in phase as shown in Figures 3a and 3b. The points along the vertical segment correspond to the Blazhko phases between 0.0 and 0.5 and the other points along the horizontal segment correspond to the second part of the Blazhko period.

In the diagram of NU Aur, the points with a magnitude fainter than 12.9 are grouped on a circle, the magnitude at maximum and (O–C) values are in phase quadrature as shown in Figures 6a and 6b. The group of points with a magnitude fainter than 12.9 are created by the non–repetitive behavior from Blazhko cycles. The full data set for NU Aur covers more than ten Blazhko cycles.

In the VY CrB graph, the points are scattered along a curve with a slope of about 135 degrees. The magnitude at maximum and (O–C) curves are in phase opposition.

For CX Lyr and VY CrB, the (O–C) errors are larger when the magnitudes at maximum are at their greatest value. This is partially due to a lower SNR but mainly because the light curve at maximum is flatter, which leads to a less precise maximum measurement.

## 6. Conclusions

This study indicates that regular observations over several seasons or years by amateurs can lead to the characterization of the Blazhko effect of RR Lyr stars: this is one of the main objectives of the professional–amateur program "GEOS RR Lyr Survey." These results should encourage amateurs to join in measurement campaigns.

The measurement of RR Lyrae stars having a strong Blazhko effect highlights the fact that this effect is not standard from one star to another, as satellite-based observations (CoRot and Kepler) have shown. Each star has a particular behavior and it may not repeat exactly from one cycle to another. A complete astrophysical model of Blazhko effect for RRab stars should be able to explain these behavior differences.

## 7. Acknowledgements

France. The authors acknowledge AAVSO Director Arne Henden and the AAVSO for the use of the AAVSONet telescopes at Sonoita (Arizona) and Cloudcroft (New Mexico). They would also like to thank G. Maintz and A. Paschke for their contributions to the GEOS database.

The authors wish to recognize their affiliation with the following organizations or institutions: Pierre de Ponthière—AAVSO, Groupe Européen d'Observations Stellaires, France (GEOS); Jean–François Le Borgne—GEOS, Université de Toulouse, France; F. Fumagalli—GEOS; Franz–Josef Hambsch—AAVSO, Bundesdeutsche Arbeitsgemeinschaft für Veränderliche Sterne e.V., Germany (BAV), GEOS, Vereniging Voor Sterrenkunde, Belgium (VVS); Tom Krajci—AAVSO; Kenneth Menzies—AAVSO, GEOS; Marco Nobile—GEOS; Richard Sabo—AAVSO, GEOS.

**References**

Agerer, F., and Huebscher, J. 2002, *Inf. Bull. Var. Stars*, No. 5296, 1 (*BAV Mitteilungen* No. 152).

Antipin, S. V. 1996, *Inf. Bull. Var. Stars*, No. 4343, 1.

Berry, R., and Burnell, J. 2001, *The Handbook of Astronomical Image Processing*, Willmann–Bell, Richmond, VA.

de Ponthière, P. 2010, LESVEPHOTOMETRY automatic photometry software, http://www.dppobservatory.net

de Ponthière, P., Le Borgne, J. –F., and Hambsch, F. –J. 2009, *J. Amer. Assoc. Var. Star Obs.*, **37**, 117.

Diffraction Limited 2004, MAXIM DL image processing software, http://www.cyanogen.com

Groupe Européen d'Observation Stellaire (GEOS) 2011, GEOS RR Lyr Database, http://rr–lyr.ast.obs–mip.fr/

Huebscher, J., Lehmann, P. B., Monninger, G., Steinbach, H. –M., and Walter, F. 2010, *Inf. Bull. Var. Stars*, No. 5941, 1 (*BAV Mitteilungen* No. 212).

Huebscher, J., Steinbach, H. –M., and Walter, F. 2008, *Inf. Bull. Var. Stars*, No. 5830, 1 (*BAV Mitteilungen* No. 193).

Huebscher, J., Steinbach, H. –M., and Walter, F. 2009, *Inf. Bull. Var. Stars*, No. 5874, 1 (*BAV Mitteilungen* No. 201).

Le Borgne, J. F., *et al.* 2007, *Astron. Astrophys.*, **476**, 307.

Poretti, E., *et al.* 2008, *Mem. Soc. Astron. Ital.*, **79**, 471.

Reinsch, C. H. 1967, *Numer. Math.*, **10**, 177.

Samolyk, G. 2011, *J. Amer. Assoc. Var. Star Obs.*, **39**, 23.

Samus, N. N., *et al.* 2011, *General Catalogue of Variable Stars* (GCVS database, Version 2011Jan), http://www.sai.msu.su/gcvs/gcvs/index.htm

Space Telescope Science Institute 2001, *The Guide Star Catalog*, Version 2.2 (VizieR On–line Data Catalog: I/271), STScI, Baltimore.

Szabó, R., *et al.* 2010, *Mon. Not. Roy. Astron. Soc.*, **409**, 1244 (http://arxiv.org/abs/1007.3404).

Vanmunster, T. 2007, peranso period analysis software, http://www.cbabelgium.com and http://www.peranso.com

Zacharias, N., Monet, D., Levine, S., Urban, S., Gaume, R., and Wycoff, G. 2011, The Naval Observatory Merged Astrometric Dataset (NOMAD), http://www.usno.navy.mil/USNO/astrometry/optical–IR–prod/nomad/

Table 1. Comparison stars for CX Lyr.

| Identification R.A. (2000) h m s | Dec. (2000) ° ′ ″ | B | V | B–V | DPP | Hambsch | Sabo |
|---|---|---|---|---|---|---|---|
| GSC 2121–2818 | | | | | | | |
| 18 51 51.48 | +28 49 08.15 | 11.129 | 10.548 | 0.581 | C1 | | |
| GSC 2121–2053 | | | | | | | |
| 18 51 37.00 | +28 51 16.32 | 11.054 | 10.565 | 0.489 | C2 | | C2 |
| GSC 2121–1980 | | | | | | | |
| 18 51 14.25 | +28 43 37.01 | 13.29 | 12.74 | 0.55 | C3 | C1 | C1 |
| GSC 2121–2842 | | | | | | | |
| 18 51 07.01 | +28 45 12.52 | 13.8 | 13.1 | 0.7 | | C2 | |

Table 2. List of measured maxima of CX Lyr.

| Maximum HJD | Error | O–C (day) | E | Magnitude | Error | Filter | Observer |
|---|---|---|---|---|---|---|---|
| 2454238.4240 | 0.0050 | –0.0139 | –712 | — | — | — | F. Fumagalli |
| 2454278.5276 | 0.0014 | 0.0004 | –647 | — | — | — | F. Fumagalli |
| 2454280.3802 | 0.0015 | 0.0027 | –644 | — | — | — | F. Fumagalli |
| 2454362.4056 | 0.0014 | –0.0007 | –511 | — | — | — | G. Maintz |
| 2454637.4929 | 0.0026 | 0.0125 | –65 | 12.329 | 0.050 | V | F.-J. Hambsch |
| 2454661.5051 | 0.0020 | –0.0289 | –26 | 12.308 | 0.050 | V | F.-J. Hambsch |
| 2454677.5688 | 0.0018 | –0.0009 | 0 | 12.152 | 0.006 | V | P. de Ponthière |
| 2454685.5930 | 0.0020 | 0.0055 | 13 | 12.062 | 0.026 | V | P. de Ponthière |
| 2454692.3815 | 0.0013 | 0.0096 | 24 | 12.051 | 0.005 | V | P. de Ponthière |
| 2454708.4197 | 0.0019 | 0.0121 | 50 | 12.235 | 0.006 | V | P. de Ponthière |
| 2454711.5050 | 0.0040 | 0.0136 | 55 | 12.237 | 0.040 | V | P. de Ponthière |
| 2454719.5020 | 0.0050 | –0.0072 | 68 | 12.297 | 0.026 | V | P. de Ponthière |
| 2454724.4300 | 0.0040 | –0.0133 | 76 | 12.331 | 0.007 | V | P. de Ponthière |
| 2454729.3630 | 0.0030 | –0.0144 | 84 | 12.324 | 0.006 | V | P. de Ponthière |
| 2454750.3518 | 0.0016 | 0.0047 | 118 | 12.134 | 0.012 | V | P. de Ponthière |
| 2454758.3736 | 0.0012 | 0.0086 | 131 | 12.156 | 0.020 | V | P. de Ponthière |
| 2454774.4100 | 0.0030 | 0.0093 | 157 | 12.293 | 0.020 | V | P. de Ponthière |
| 2454782.3940 | 0.0030 | –0.0246 | 170 | 12.363 | 0.022 | V | P. de Ponthière |
| 2454983.4954 | 0.0024 | 0.0137 | 496 | 12.333 | 0.008 | V | P. de Ponthière |
| 2455041.4761 | 0.0032 | 0.0191 | 590 | 12.158 | 0.008 | V | P. de Ponthière |

Table 2. List of measured maxima of CX Lyr, cont.

| Maximum HJD | Error | O–C (day) | E | Magnitude | Error | Filter | Observer |
|---|---|---|---|---|---|---|---|
| 2455046.4025 | 0.0025 | 0.0115 | 598 | 12.296 | 0.013 | V | F.-J. Hambsch |
| 2455049.4886 | 0.0035 | 0.0138 | 603 | 12.294 | 0.010 | V | P. de Ponthière |
| 2455052.5695 | 0.0085 | 0.0109 | 608 | 12.295 | 0.015 | V | P. de Ponthière |
| 2455057.4596 | 0.0060 | –0.0331 | 616 | — | — | V | G. Maintz |
| 2455060.5485 | 0.0084 | –0.0279 | 621 | 12.371 | 0.010 | V | P. de Ponthière |
| 2455062.3950 | 0.0010 | –0.0317 | 624 | — | — | V | G. Maintz |
| 2455062.3977 | 0.0032 | –0.0290 | 624 | 12.432 | 0.009 | V | F.-J. Hambsch |
| 2455094.5062 | 0.0025 | 0.0081 | 676 | 12.065 | 0.015 | V | P. de Ponthière |
| 2455096.3587 | 0.0015 | 0.0103 | 679 | 12.042 | 0.008 | V | P. de Ponthière |
| 2455295.5698 | 0.0015 | 0.0085 | 1002 | 12.087 | 0.011 | V | P. de Ponthière |
| 2455303.5902 | 0.0012 | 0.0111 | 1015 | 12.021 | 0.010 | V | P. de Ponthière |
| 2455308.5270 | 0.0020 | 0.0138 | 1023 | 12.024 | 0.011 | V | P. de Ponthière |
| 2455311.6080 | 0.0025 | 0.0110 | 1028 | 12.073 | 0.020 | V | P. de Ponthière |
| 2455312.8464 | 0.0018 | 0.0159 | 1030 | 12.162 | 0.012 | V | F.-J. Hambsch |
| 2455320.8601 | 0.0019 | 0.0117 | 1043 | 12.307 | 0.011 | V | F.-J. Hambsch |
| 2455333.7745 | 0.0040 | –0.0258 | 1064 | 12.397 | 0.012 | V | F.-J. Hambsch |
| 2455353.5350 | 0.0025 | –0.0015 | 1096 | 12.303 | 0.012 | V | P. de Ponthière |
| 2455363.4131 | 0.0025 | 0.0084 | 1112 | 12.122 | 0.030 | V | P. de Ponthière |
| 2455369.5828 | 0.0020 | 0.0106 | 1122 | 12.050 | 0.010 | V | P. de Ponthière |
| 2455374.5167 | 0.0020 | 0.0104 | 1130 | 12.030 | 0.010 | V | P. de Ponthière |
| 2455379.4518 | 0.0020 | 0.0114 | 1138 | 12.056 | 0.009 | V | P. de Ponthière |
| 2455382.5342 | 0.0015 | 0.0100 | 1143 | 12.121 | 0.009 | V | P. de Ponthière |
| 2455387.4662 | 0.0023 | 0.0080 | 1151 | 12.188 | 0.010 | V | P. de Ponthière |
| 2455392.3970 | 0.0030 | 0.0047 | 1159 | 12.286 | 0.010 | V | P. de Ponthière |
| 2455395.4753 | 0.0040 | –0.0008 | 1164 | 12.337 | 0.009 | V | P. de Ponthière |
| 2455398.5497 | 0.0074 | –0.0102 | 1169 | 12.364 | 0.009 | V | P. de Ponthière |
| 2455408.3959 | 0.0035 | –0.0321 | 1185 | 12.325 | 0.010 | V | P. de Ponthière |
| 2455429.4044 | 0.0025 | 0.0066 | 1219 | 12.189 | 0.011 | V | P. de Ponthière |
| 2455440.5115 | 0.0017 | 0.0121 | 1237 | 11.988 | 0.009 | V | P. de Ponthière |
| 2455442.3601 | 0.0020 | 0.0104 | 1240 | 11.980 | 0.012 | V | P. de Ponthière |
| 2455445.4485 | 0.0015 | 0.0150 | 1245 | 12.009 | 0.008 | V | P. de Ponthière |
| 2455461.4862 | 0.0040 | 0.0170 | 1271 | 12.284 | 0.011 | V | P. de Ponthière |
| 2455470.7015 | 0.0065 | –0.0191 | 1286 | 12.427 | 0.011 | V | F.-J. Hambsch |
| 2455479.3298 | 0.0047 | –0.0254 | 1300 | 12.371 | 0.010 | V | P. de Ponthière |
| 2455492.3090 | 0.0075 | 0.0019 | 1321 | 12.347 | 0.018 | V | P. de Ponthière |
| 2455649.5879 | 0.0020 | 0.0075 | 1576 | 12.053 | 0.009 | V | P. de Ponthière |
| 2455670.5422 | 0.0044 | –0.0080 | 1610 | 12.343 | 0.010 | V | P. de Ponthière |
| 2455713.7302 | 0.0013 | 0.0069 | 1680 | 12.025 | 0.006 | V | K. Menzies |
| 2455745.7679 | 0.0020 | –0.0268 | 1732 | 12.464 | 0.004 | V | R. Sabo |
| 2455746.3925 | 0.0026 | –0.0189 | 1733 | 12.401 | 0.013 | V | P. de Ponthière |
| 2455807.4625 | 0.0034 | –0.0080 | 1832 | 12.366 | 0.010 | V | P. de Ponthière |

Table 3. Comparison stars for NU Aur.

| Identification R.A. (2000) h m s | Dec. (2000) ° ' " | B | V | B–V | DPP | Hambsch | Menzies | Sabo |
|---|---|---|---|---|---|---|---|---|
| GSC 1857–1453 | | | | | | | | |
| 05 08 58.82 | +28 42 08 | 13.32 | 12.53 | 0.795 | C1 | C1 | C1 | C1 |
| GSC 1857–1288 | | | | | | | | |
| 05 08 59.15 | +28 43 20.2 | | 13.67 | | | C2 | | |
| GSC 1857–1288 | | | | | | | | |
| 05 08 59.17 | +28 43 20.3 | 14.46 | 13.77 | 0.69 | C2 | | C2 | C2 |
| GSC 1857–938 | | | | | | | | |
| 05 08 34.00 | +28 45 07.6 | 13.322 | 12.326 | 0.996 | C3 | C3 | C3 | C3 |

Table 4. List of measured maxima of NU Aur.

| Maximum HJD | Error | O–C (day) | E | Magnitude | Error | Filter | Observer |
|---|---|---|---|---|---|---|---|
| 2454081.4445 | 0.0100 | 0.0204 | –1244 | — | — | — | J.-M. Llapasset |
| 2454083.5942 | 0.0005 | 0.0125 | –1240 | — | — | — | J.-M. Llapasset |
| 2454088.4454 | 0.0010 | 0.0089 | –1231 | — | — | — | J.-M. Llapasset |
| 2454089.5225 | 0.0005 | 0.0072 | –1229 | — | — | — | J.-M. Llapasset |
| 2454090.6033 | 0.0005 | 0.0092 | –1227 | — | — | — | J.-M. Llapasset |
| 2454091.6733 | 0.0010 | 0.0003 | –1225 | — | — | — | J.-M. Llapasset |
| 2454096.5397 | 0.0005 | 0.0120 | –1216 | — | — | — | J.-M. Llapasset |
| 2454100.3214 | 0.0005 | 0.0178 | –1209 | — | — | — | J.-M. Llapasset |
| 2454107.3333 | 0.0010 | 0.0173 | –1196 | — | — | — | J.-M. Llapasset |
| 2454114.3359 | 0.0005 | 0.0075 | –1183 | — | — | — | J.-M. Llapasset |
| 2454121.3493 | 0.0005 | 0.0085 | –1170 | — | — | — | J.-M. Llapasset |
| 2454128.3628 | 0.0010 | 0.0096 | –1157 | — | — | — | J.-M. Llapasset |
| 2454135.3748 | 0.0005 | 0.0092 | –1144 | — | — | — | J.-M. Llapasset |
| 2454456.3178 | 0.0004 | –0.0001 | –549 | — | — | — | G. Maintz |
| 2454752.4570 | 0.0040 | 0.0000 | 0 | 13.042 | 0.011 | C | P. de Ponthière |
| 2454759.4820 | 0.0030 | 0.0126 | 13 | 13.028 | 0.008 | C | P. de Ponthière |
| 2454774.5770 | 0.0020 | 0.0040 | 41 | 12.899 | 0.012 | C | P. de Ponthière |
| 2454787.5310 | 0.0040 | 0.0120 | 65 | 12.860 | 0.007 | C | P. de Ponthière |
| 2454801.0080 | 0.0030 | 0.0036 | 90 | 12.737 | 0.005 | V | P. de Ponthière |
| 2454804.7820 | 0.0040 | 0.0017 | 97 | 12.875 | 0.005 | V | P. de Ponthière |
| 2454808.0200 | 0.0030 | 0.0032 | 103 | 12.845 | 0.005 | V | P. de Ponthière |
| 2454827.4315 | 0.0018 | –0.0043 | 139 | 12.854 | 0.007 | C | P. de Ponthière |
| 2454828.5107 | 0.0017 | –0.0039 | 141 | 12.838 | 0.007 | C | P. de Ponthière |
| 2454829.5870 | 0.0030 | –0.0064 | 143 | 12.846 | 0.009 | C | P. de Ponthière |
| 2454829.5900 | 0.0020 | –0.0034 | 143 | 12.852 | 0.009 | V | P. de Ponthière |
| 2454830.6570 | 0.0030 | –0.0153 | 145 | 12.878 | 0.009 | V | P. de Ponthière |

Table 4. List of measured maxima of NU Aur, cont.

| Maximum HJD | Error | O–C (day) | E | Magnitude | Error | Filter | Observer |
|---|---|---|---|---|---|---|---|
| 2454830.6610 | 0.0030 | –0.0113 | 145 | 12.902 | 0.014 | V | P. de Ponthière |
| 2454831.7451 | 0.0017 | –0.0060 | 147 | 12.849 | 0.008 | V | P. de Ponthière |
| 2454832.8202 | 0.0016 | –0.0097 | 149 | 12.875 | 0.008 | V | P. de Ponthière |
| 2454833.9020 | 0.0030 | –0.0067 | 151 | 12.891 | 0.009 | V | P. de Ponthière |
| 2454838.7576 | 0.0014 | –0.0059 | 160 | 12.931 | 0.011 | V | P. de Ponthière |
| 2454841.4540 | 0.0050 | –0.0066 | 165 | 12.927 | 0.023 | V | P. de Ponthière |
| 2454843.6114 | 0.0018 | –0.0068 | 169 | 12.936 | 0.013 | V | P. de Ponthière |
| 2454844.6863 | 0.0012 | –0.0108 | 171 | 12.902 | 0.012 | V | P. de Ponthière |
| 2454845.7661 | 0.0015 | –0.0098 | 173 | 12.900 | 0.011 | V | P. de Ponthière |
| 2454846.3070 | 0.0020 | –0.0083 | 174 | — | — | — | J.-M. Llapasset |
| 2454846.8450 | 0.0020 | –0.0097 | 175 | 12.898 | 0.010 | V | P. de Ponthière |
| 2454850.6150 | 0.0020 | –0.0156 | 182 | 12.904 | 0.010 | V | P. de Ponthière |
| 2454851.6940 | 0.0040 | –0.0155 | 184 | 12.946 | 0.009 | V | P. de Ponthière |
| 2454852.7740 | 0.0020 | –0.0143 | 186 | 12.932 | 0.006 | V | P. de Ponthière |
| 2454857.6330 | 0.0040 | –0.0100 | 195 | 13.018 | 0.008 | V | P. de Ponthière |
| 2454860.3380 | 0.0050 | –0.0021 | 200 | 13.021 | 0.016 | V | P. de Ponthière |
| 2454861.4150 | 0.0040 | –0.0039 | 202 | 13.040 | 0.018 | V | P. de Ponthière |
| 2454862.4960 | 0.0040 | –0.0018 | 204 | 13.034 | 0.017 | V | P. de Ponthière |
| 2454884.6190 | 0.0018 | 0.0051 | 245 | 13.007 | 0.008 | V | P. de Ponthière |
| 2454887.3170 | 0.0020 | 0.0061 | 250 | — | — | — | J.-M. Llapasset |
| 2454888.3980 | 0.0020 | 0.0083 | 252 | — | — | — | J.-M. Llapasset |
| 2454891.6396 | 0.0030 | 0.0133 | 258 | 13.017 | 0.011 | V | P. de Ponthière |
| 2455100.9247 | 0.0031 | 0.0053 | 646 | — | — | V | R. Sabo |
| 2455106.8653 | 0.0036 | 0.0123 | 657 | 12.995 | 0.019 | V | F.-J. Hambsch |
| 2455113.8751 | 0.0028 | 0.0097 | 670 | 12.937 | 0.017 | V | F.-J. Hambsch |
| 2455114.9502 | 0.0040 | 0.0060 | 672 | 12.961 | 0.036 | V | F.-J. Hambsch |
| 2455119.8273 | 0.0035 | 0.0283 | 681 | 12.971 | 0.019 | V | F.-J. Hambsch |
| 2455120.8963 | 0.0029 | 0.0185 | 683 | 12.923 | 0.011 | V | F.-J. Hambsch |
| 2455121.9780 | 0.0023 | 0.0214 | 685 | 12.889 | 0.009 | V | F.-J. Hambsch |
| 2455122.5074 | 0.0030 | 0.0113 | 686 | 12.878 | 0.010 | V | F.-J. Hambsch |
| 2455127.9098 | 0.0034 | 0.0196 | 696 | 12.797 | 0.019 | V | F.-J. Hambsch |
| 2455128.9865 | 0.0039 | 0.0175 | 698 | 12.783 | 0.003 | V | F.-J. Hambsch |
| 2455135.9890 | 0.0120 | 0.0076 | 711 | 12.674 | 0.008 | V | F.-J. Hambsch |
| 2455155.4042 | 0.0019 | 0.0038 | 747 | 12.703 | 0.009 | V | F.-J. Hambsch |
| 2455175.3210 | 0.0020 | –0.0378 | 784 | — | — | — | M. Nobile |
| 2455182.9060 | 0.0032 | –0.0046 | 798 | 12.890 | 0.013 | V | F.-J. Hambsch |
| 2455189.3650 | 0.0040 | –0.0186 | 810 | 12.867 | 0.015 | V | F.-J. Hambsch |
| 2455208.8035 | 0.0030 | 0.0010 | 846 | 13.014 | 0.008 | V | F.-J. Hambsch |
| 2455241.7199 | 0.0025 | 0.0130 | 907 | 12.963 | 0.010 | V | F.-J. Hambsch |

Table 4. List of measured maxima of NU Aur, cont.

| Maximum HJD | Error | O–C (day) | E | Magnitude | Error | Filter | Observer |
|---|---|---|---|---|---|---|---|
| 2455247.6585 | 0.0030 | 0.0180 | 918 | 12.862 | 0.009 | V | F.-J. Hambsch |
| 2455261.6812 | 0.0048 | 0.0159 | 944 | 12.898 | 0.009 | V | F.-J. Hambsch |
| 2455479.6006 | 0.0025 | 0.0115 | 1348 | 12.713 | 0.013 | V | P. de Ponthière |
| 2455481.7610 | 0.0026 | 0.0142 | 1352 | 12.741 | 0.010 | V | F.-J. Hambsch |
| 2455492.5516 | 0.0034 | 0.0165 | 1372 | 12.717 | 0.026 | V | P. de Ponthière |
| 2455511.9491 | 0.0020 | –0.0050 | 1408 | 12.686 | 0.010 | V | F.-J. Hambsch |
| 2455528.6650 | 0.0024 | –0.0110 | 1439 | 12.781 | 0.012 | V | F.-J. Hambsch |
| 2455531.8989 | 0.0031 | –0.0135 | 1445 | 12.847 | 0.010 | V | F.-J. Hambsch |
| 2455538.9143 | 0.0040 | –0.0105 | 1458 | 13.014 | 0.010 | V | F.-J. Hambsch |
| 2455540.5315 | 0.0075 | –0.0116 | 1461 | 13.044 | 0.013 | V | K. Menzies |
| 2455542.6905 | 0.0041 | –0.0103 | 1465 | 13.026 | 0.013 | V | F.-J. Hambsch |
| 2455543.7760 | 0.0048 | –0.0036 | 1467 | 13.062 | 0.011 | V | F.-J. Hambsch |
| 2455545.9362 | 0.0042 | –0.0010 | 1471 | 13.016 | 0.011 | V | F.-J. Hambsch |
| 2455548.6347 | 0.0036 | 0.0004 | 1476 | 13.066 | 0.008 | V | R. Sabo |
| 2455554.5716 | 0.0050 | 0.0037 | 1487 | 13.115 | 0.013 | V | K. Menzies |
| 2455555.6427 | 0.0035 | –0.0040 | 1489 | 13.031 | 0.011 | V | K. Menzies |
| 2455571.3082 | 0.0060 | 0.0184 | 1518 | 13.057 | 0.016 | V | P. de Ponthière |
| 2455572.3921 | 0.0038 | 0.0235 | 1520 | 13.063 | 0.010 | V | F.-J. Hambsch |
| 2455575.6360 | 0.0060 | 0.0309 | 1526 | 13.045 | 0.001 | V | F.-J. Hambsch |
| 2455583.7160 | 0.0031 | 0.0197 | 1541 | 12.941 | 0.012 | V | F.-J. Hambsch |
| 2455589.6450 | 0.0047 | 0.0151 | 1552 | 12.919 | 0.012 | V | K. Menzies |
| 2455627.3941 | 0.0030 | 0.0051 | 1622 | 12.767 | 0.015 | V | P. de Ponthière |
| 2455627.3945 | 0.0016 | 0.0055 | 1622 | 12.775 | 0.006 | V | F.-J. Hambsch |
| 2455640.3375 | 0.0040 | 0.0025 | 1646 | 12.787 | 0.015 | V | P. de Ponthière |

Table 5. Comparison stars for VY CrB.

| Identification R.A. (2000) h m s | Dec. (2000) ° ′ ″ | B | V | B–V | DPP | Hambsch | Menzies | Sabo |
|---|---|---|---|---|---|---|---|---|
| GSC 2576–1883 | | | | | | | | |
| 16 06 45.0 | +33 19 35.756 | 12.265 | 11.638 | 0.627 | C1 | | C1 | C1 |
| GSC 2576–1372 | | | | | | | | |
| 16 05 53.7 | +33 20 17.166 | 14.97 | 13.71 | 1.26 | C2 | C1 | C2 | |
| GSC 2576–740 | | | | | | | | |
| 16 06 13.7 | +33 19 05.222 | 14.36 | 13.58 | 0.78 | C3 | C2 | C3 | C2 |
| GSC 2576–1021 | | | | | | | | |
| 16 06 05.5 | +33 25 00.460 | 16.77 | 14.65 | 2.12 | C4 | C3 | C4 | |

Table 6. List of measured maxima of VY CrB.

| Maximum HJD | Error | O–C (day) | E | Magnitude | Error | Filter | Observer |
|---|---|---|---|---|---|---|---|
| 2451615.5930 | 0.005 | –0.0153 | –7964 | — | — | — | A. Paschke |
| 2454215.5169 | 0.002 | 0.0089 | –2348 | — | — | — | J.-M. Llapasset |
| 2454216.4421 | 0.002 | 0.0082 | –2346 | — | — | — | J.-M. Llapasset |
| 2455302.4926 | 0.0013 | –0.0105 | 0 | 13.480 | 0.009 | C | P. de Ponthière |
| 2455303.4196 | 0.0018 | –0.0094 | 2 | 13.510 | 0.012 | C | P. de Ponthière |
| 2455309.4562 | 0.0031 | 0.0088 | 15 | 13.662 | 0.009 | C | P. de Ponthière |
| 2455310.3791 | 0.0023 | 0.0059 | 17 | 13.702 | 0.011 | C | P. de Ponthière |
| 2455321.4822 | 0.0021 | –0.0017 | 41 | 13.548 | 0.008 | C | P. de Ponthière |
| 2455352.5027 | 0.0024 | 0.0015 | 108 | 13.602 | 0.008 | C | P. de Ponthière |
| 2455364.5240 | 0.0014 | –0.0138 | 134 | 13.376 | 0.009 | C | P. de Ponthière |
| 2455370.5535 | 0.003 | –0.0026 | 147 | 13.557 | 0.013 | C | P. de Ponthière |
| 2455371.4812 | 0.0024 | –0.0008 | 149 | 13.570 | 0.009 | C | P. de Ponthière |
| 2455377.5180 | 0.008 | 0.0177 | 162 | 13.733 | 0.030 | C | P. de Ponthière |
| 2455378.4431 | 0.0053 | 0.0170 | 164 | 13.719 | 0.014 | C | P. de Ponthière |
| 2455384.4469 | 0.0033 | 0.0025 | 177 | 13.614 | 0.009 | C | P. de Ponthière |
| 2455390.4495 | 0.0016 | –0.0132 | 190 | 13.352 | 0.009 | C | P. de Ponthière |
| 2455391.3773 | 0.0015 | –0.0113 | 192 | 13.343 | 0.009 | C | P. de Ponthière |
| 2455396.4673 | 0.0016 | –0.0137 | 203 | 13.343 | 0.007 | C | P. de Ponthière |
| 2455397.3939 | 0.0017 | –0.0130 | 205 | 13.372 | 0.007 | C | P. de Ponthière |
| 2455410.3855 | 0.0048 | 0.0161 | 233 | 13.737 | 0.009 | C | P. de Ponthière |
| 2455441.3929 | 0.0029 | 0.0062 | 300 | 13.743 | 0.014 | C | P. de Ponthière |
| 2455461.2817 | 0.0015 | –0.0116 | 343 | 13.347 | 0.008 | C | P. de Ponthière |
| 2455480.2825 | 0.0036 | 0.0084 | 384 | 13.676 | 0.009 | C | P. de Ponthière |
| 2455627.4899 | 0.003 | –0.0007 | 702 | 13.555 | 0.010 | C | P. de Ponthière |
| 2455640.4717 | 0.007 | 0.0186 | 730 | 13.754 | 0.011 | C | P. de Ponthière |
| 2455622.8556 | 0.0054 | –0.0056 | 692 | 13.316 | 0.058 | V | K. Menzies |
| 2455644.6183 | 0.0026 | –0.0013 | 739 | 13.581 | 0.009 | C | P. de Ponthière |
| 2455645.5423 | 0.003 | –0.0032 | 741 | 13.559 | 0.009 | C | P. de Ponthière |
| 2455646.4724 | 0.0041 | 0.0010 | 743 | 13.613 | 0.022 | C | P. de Ponthière |
| 2455602.9682 | 0.005 | 0.0137 | 649 | 13.717 | 0.015 | V | F.-J. Hambsch |
| 2455608.9772 | 0.005 | 0.0044 | 662 | 13.680 | 0.018 | V | F.-J. Hambsch |
| 2455609.9030 | 0.005 | 0.0043 | 664 | 13.661 | 0.025 | V | F.-J. Hambsch |
| 2455614.9834 | 0.0028 | –0.0077 | 675 | 13.413 | 0.015 | V | F.-J. Hambsch |
| 2455615.9087 | 0.003 | –0.0083 | 677 | 13.370 | 0.019 | V | F.-J. Hambsch |
| 2455622.8554 | 0.002 | –0.0058 | 692 | 13.294 | 0.012 | V | F.-J. Hambsch |
| 2455629.8069 | 0.004 | 0.0016 | 707 | 13.582 | 0.018 | V | F.-J. Hambsch |
| 2455640.9245 | 0.007 | 0.0085 | 731 | 13.690 | 0.024 | V | F.-J. Hambsch |
| 2455646.9268 | 0.003 | –0.0075 | 744 | 13.432 | 0.014 | V | F.-J. Hambsch |
| 2455647.8531 | 0.003 | –0.0071 | 746 | 13.382 | 0.013 | V | F.-J. Hambsch |

Table 6. List of measured maxima of VY CrB, cont.

| Maximum HJD | Error | O–C (day) | E | Magnitude | Error | Filter | Observer |
|---|---|---|---|---|---|---|---|
| 2455653.8707 | 0.0022 | –0.0078 | 759 | 13.276 | 0.013 | V | F.-J. Hambsch |
| 2455654.7969 | 0.0019 | –0.0075 | 761 | 13.276 | 0.013 | V | F.-J. Hambsch |
| 2455660.8238 | 0.003 | 0.0011 | 774 | 13.545 | 0.014 | V | F.-J. Hambsch |
| 2455665.9194 | 0.0053 | 0.0043 | 785 | 13.695 | 0.018 | V | F.-J. Hambsch |
| 2455666.8510 | 0.0044 | 0.0101 | 787 | 13.715 | 0.020 | V | F.-J. Hambsch |
| 2455671.4845 | 0.0039 | 0.0141 | 797 | 13.748 | 0.011 | C | P. de Ponthière |
| 2455672.4003 | 0.0047 | 0.0040 | 799 | 13.769 | 0.012 | C | P. de Ponthière |
| 2455714.5166 | 0.0013 | –0.0077 | 890 | 13.337 | 0.007 | C | P. de Ponthière |
| 2455715.4414 | 0.0014 | –0.0088 | 892 | 13.325 | 0.007 | C | P. de Ponthière |
| 2455760.8215 | 0.003 | 0.0027 | 990 | 13.690 | 0.011 | V | R. Sabo |
| 2455739.5326 | 0.0041 | 0.0093 | 944 | 13.608 | 0.009 | C | P. de Ponthière |
| 2455775.6294 | 0.0025 | –0.0036 | 1022 | 13.494 | 0.013 | V | K. Menzies |
| 2455784.4199 | 0.003 | –0.0091 | 1041 | 13.367 | 0.010 | C | P. de Ponthière |

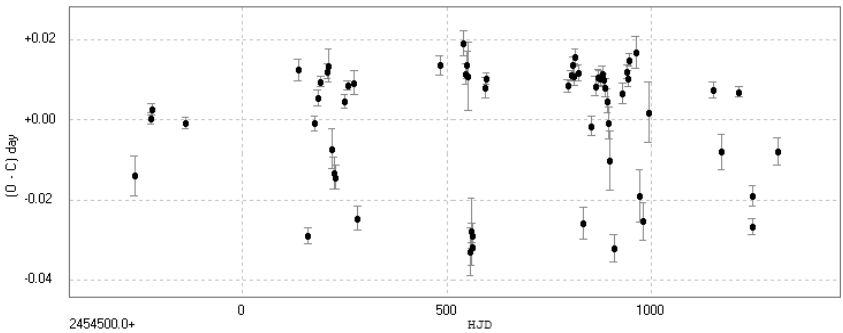

Figure 1. CX Lyr (O–C).

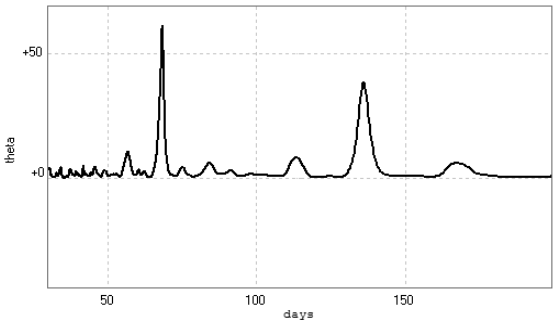

Figure 2. CX Lyr (O–C) periodogram.

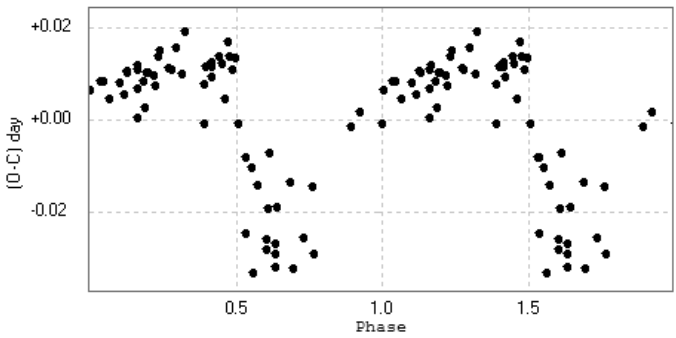

Figure 3a. CX Lyr (O–C) at maximum versus Blazhko phase.

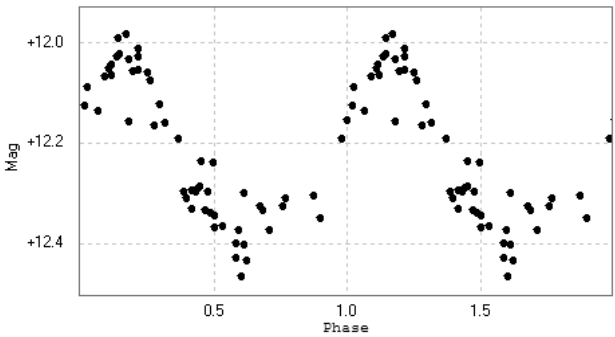

Figure 3b. CX Lyr magnitude at maximum versus Blazhko phase.

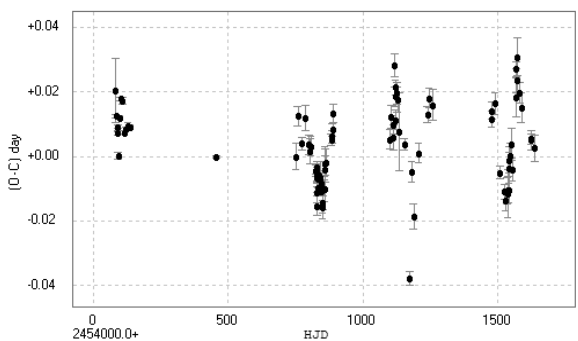

Figure 4. NU Aur (O–C).

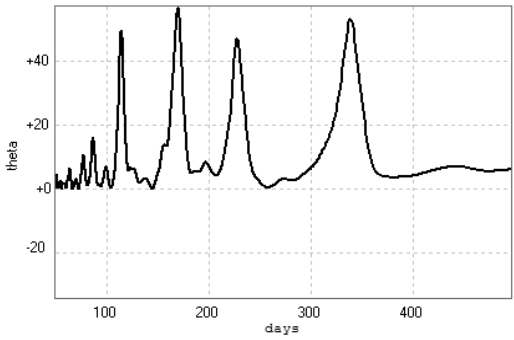

Figure 5a. NU Aur (O–C) periodogram.

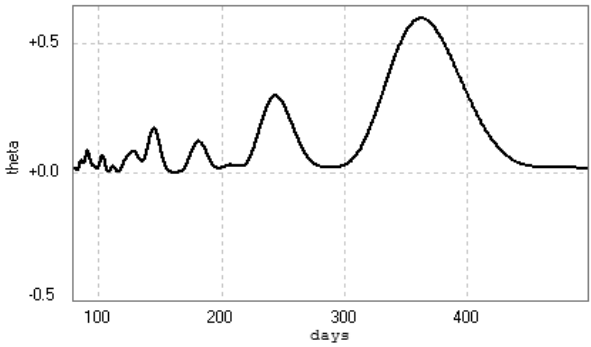

Figure 5b. NU Aur spectral window.

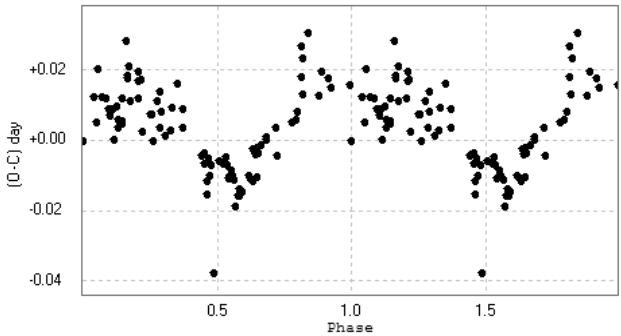

Figure 6a. NU Aur (O–C) at maximum versus Blazhko phase.

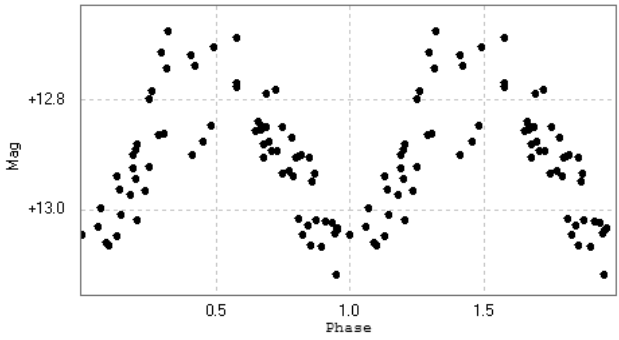

Figure 6b. NU Aur magnitude at maximum versus Blazhko phase.

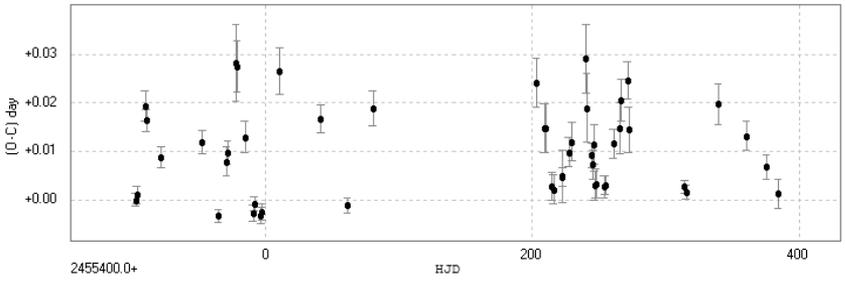

Figure 7. VY CrB (O–C).

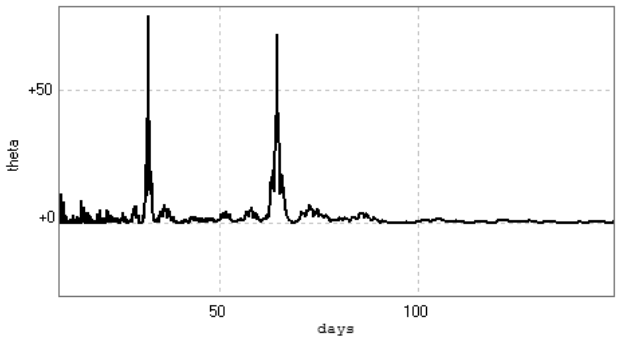

Figure 8. VY CrB (O–C) periodogram.

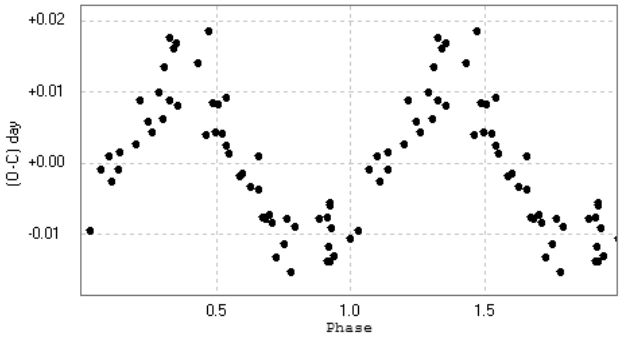

Figure 9a. VY CrB (O–C) at maximum versus Blazhko phase.

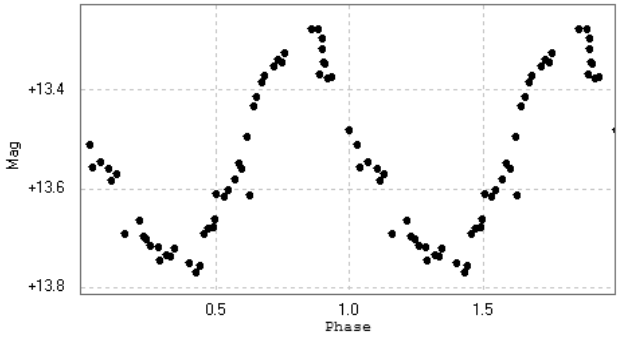

Figure 9b. VY CrB magnitude at maximum versus Blazhko phase.

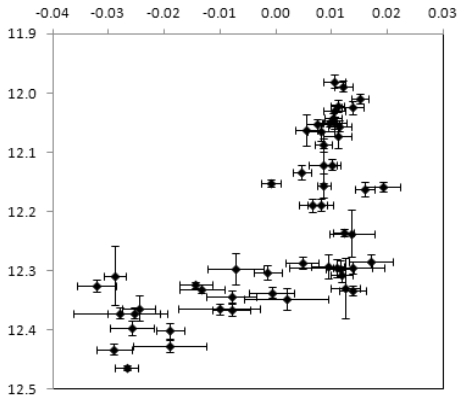

Figure 10a. CX Lyr magnitude at maximum versus (O–C) values.

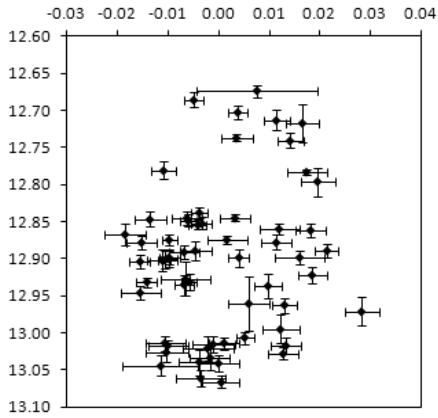

Figure 10b. NU Aur magnitude at maximum versus (O–C) values.

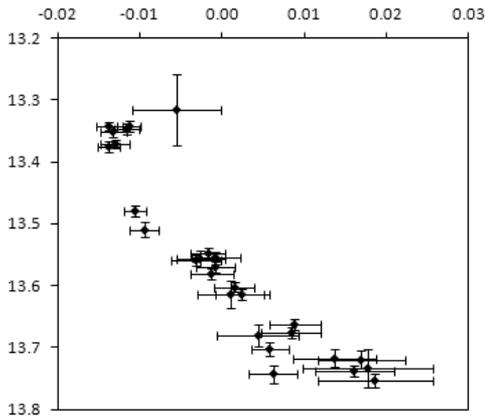

Figure 10c. VY CrB magnitude at maximum versus (O–C) values.